\documentstyle[12pt]{article}

\font\twlgot =eufm10 scaled \magstep1
\font\egtgot =eufm8
\font\sevgot =eufm7
\font\twlmsb =msbm10 scaled \magstep1
\font\egtmsb =msbm8
\font\sevmsb =msbm7

\newfam\gotfam

\textfont\gotfam\twlgot
\scriptfont\gotfam\egtgot
\scriptscriptfont\gotfam\sevgot

\newfam\msbfam
\textfont\msbfam\twlmsb
\scriptfont\msbfam\egtmsb
\scriptscriptfont\msbfam\sevmsb
\def\Bbb{\protect\pBbb}
\def\pBbb{\relax\ifmmode\expandafter\Bb\else\typeout{You cann't use
Bbb in text mode}\fi}
\def\Bb #1{{\fam\msbfam\relax#1}}

\def\thebibliography#1{\bigskip\section*{}\bigskip\list
{$^{\arabic{enumi}}$}{\settowidth\labelwidth{#1}\leftmargin\labelwidth
\advance\leftmargin\labelsep
\usecounter{enumi}}
\def\newblock{\hskip .11em plus .33em minus .07em}
\sloppy\clubpenalty4000\widowpenalty4000
\sfcode`\.=1000\relax}

\let\Large=\large

\def\op#1{\mathop{\fam0 #1}\limits}

\newcommand{\id}{{\rm Id\,}}

\newcommand{\beq}{\begin{equation}}
\newcommand{\eeq}{\end{equation}}
\newcommand{\ben}{\begin{eqnarray}}
\newcommand{\een}{\end{eqnarray}}
\newcommand{\be}{\begin{eqnarray*}}
\newcommand{\ee}{\end{eqnarray*}}
\newcommand{\bea}{\begin{eqalph}}
\newcommand{\eea}{\end{eqalph}}
\newcommand{\cA}{{\cal A}}

\newcommand{\cD}{{\cal D}}

\newcommand{\cH}{{\cal H}}
\newcommand{\cF}{{\cal F}}

\newcommand{\bL}{{\bf L}}

\newcommand{\bC}{{\bf C}}

\newcommand{\al}{\alpha}

\newcommand{\bt}{\beta}
\newcommand{\dl}{\delta}
\newcommand{\la}{\lambda}
\newcommand{\La}{\Lambda}
\newcommand{\f}{\phi}

\newcommand{\Om}{\Omega}
\newcommand{\m}{\mu}

\newcommand{\g}{\gamma}

\newcommand{\vt}{\vartheta}
\newcommand{\vf}{\varphi}

\newcommand{\lng}{\langle}
\newcommand{\rng}{\rangle}

\newcommand{\si}{\sigma}

\newcommand{\w}{\wedge}

\newcommand{\wh}{\widehat}
\newcommand{\ol}{\overline}
\newcommand{\dr}{\partial}

\newcommand{\ot}{\otimes}

\newcommand{\ve}{\varepsilon}

\let\ssection=\section
\renewcommand{\section}{\setcounter{equation}{0}\ssection}

\newcounter{eqalph}
\newcounter{equationa}
\newcounter{remark}
\newcounter{example}
\newcounter{theorem}
\newcounter{proposition}
\newcounter{lemma}
\newcounter{corollary}
\newcounter{definition}
\newenvironment{eqalph}{\stepcounter{equation}
\setcounter{equationa}{\value{equation}}
\setcounter{equation}{0}

\begin{eqnarray}}{\end{eqnarray}\setcounter{equation}{\value{equationa}}}

\def\therexample{\arabic{remark}}

\def\thedefinition{\arabic{definition}}

\newenvironment{proof}{\medskip\noindent
{\it Proof:}}{\medskip}

\newenvironment{prop}{\refstepcounter{definition} \medskip
\noindent{\it Proposition \thedefinition:}}{\medskip}

\newcommand{\mar}[1]{}

\begin{document}
\hbox{}

{\parindent=0pt

{\Large \bf Geometric quantization of time-dependent
completely integrable 
Hamiltonian systems}

\bigskip

{\sc E.Fiorani}\footnote{Electronic mail: fiorani@mat.unimi.it}

{\sl Dipartimento di Matematica "F.Enriques", Universit\'a di Milano, 
20133, Milano, Italy}

{\sc G.Giachetta}\footnote{Electronic mail: giachetta@campus.unicam.it},

{\sl Dipartimento di Matematica e Fisica, Universit\'a di Camerino,
62032 Camerino (MC), Italy}

\medskip

{\sc G. Sardanashvily}\footnote{Electronic mail:
sard@grav.phys.msu.su; URL: http://webcenter.ru/$\sim$sardan/}

{\sl Department of Theoretical Physics,
Moscow State University, 117234 Moscow, Russia}

\bigskip
A time-dependent completely
integrable Hamiltonian system is 
quantized with respect to time-dependent action-angle variables 
near an instantly compact regular invariant manifold. Its Hamiltonian
depends only on action variables, and has a
time-independent countable energy spectrum. 
}
\bigskip
\bigskip

\noindent
{\bf I. INTRODUCTION}
\bigskip

A time-dependent Hamiltonian system of $m$ degrees of freedom is called
a completely integrable system (henceforth CIS)
if it admits $m$ independent first integrals 
in involution.
Choosing appropriate dynamic variables, 
one may hope to quantize a time-dependent CIS
so that its quantum Hamiltonian and 
first integral operators possess time-independent spectra.$^1$
Time-dependent action-angle variables introduced below are of this type.
Written relative to these variables, 
a Hamiltonian of a time-dependent CIS is a function only of the action coordinates.
It follows that, if time-dependent 
action-angle coordinates hold fixed, a time-dependent
CIS can be quantized just as an autonomous one, and its energy spectrum
is time-independent.

In order to introduce time-dependent action-angle variables,
we use the fact that a
time-dependent CIS of $m$ degrees of freedom 
can be extended to an 
autonomous one of $m+1$ degrees of
freedom where the time is regarded as a dynamic variable.$^{2-4}$  
By virtue of the 
classical Arnold--Liouville theorem,$^{5,6}$ an autonomous CIS admits
the action-angle coordinates around a regular connected 
compact invariant manifold.
The problem is that 
invariant manifolds of a time-dependent CIS
are not compact because of the time axis. Therefore, we first
generalize the above mentioned theorem to noncompact invariant
manifolds.
Then we show that, if a regular connected invariant manifold $N$
of a time-dependent CIS 
is compact at each instant, it admits
an open neighbourhood in the ambient momentum phase space which is
isomorphic to the product
\mar{z48}\beq
W= \Bbb R\times T^m \times V\label{z48}
\eeq
of the time axis $\Bbb R$, an $m$-dimensional torus $T^m$
and an open domain $V\subset \Bbb R^m$. This product
is equipped with the coordinates 
\mar{ww6}\beq
(t,\f^i,I_i), \qquad i=1,\ldots,m, \label{ww6}
\eeq
where $t$ is the Cartesian coordinate on $\Bbb R$ and
$\f^i$ are cyclic coordinates on $T^m$. 
Written with respect to these coordinates, a Hamiltonian
and first integrals of a time-dependent CIS are functions only
of action coordinates $I_i$. The corresponding Hamilton equation 
on $W$ reads
\be
\dot I_i=0, \qquad
\dot\f^i=\dr^i\cH(I_j). 
\ee

A glance at this equation shows
shows that, given 
action-angle coordinates (\ref{ww6}), a time-dependent
CIS can be seen as an autonomous CIS
on the symplectic annulus 
\mar{00}\beq
P=V\times T^m, \label{00}
\eeq
equipped with the action-angle coordinates $(\f^i,I_i)$ and 
provided with the symplectic form 
\mar{ci1}\beq
\Om_P=dI_i\w d\f^i. \label{ci1}
\eeq
Therefore, we can
quantize a time-dependent CIS with respect to action-angle
variables similarly to that of an autonomous CIS. Of course,
the choice of time-dependent action-angle coordinates by no
means is unique. They are changed by
canonical transformations. Therefore, we employ the
geometric quantization technique$^{7-9}$ which remains equivalent 
under such kind transformations.
At the same time, geometric quantization
essentially depends on the choice of polarization.$^{10,11}$

Geometric quantization of
an autonomous CIS has been studied 
with respect to polarization spanned by Hamiltonian vector fields of
first integrals.$^{12}$ In fact, the Simms quantization of
the harmonic oscillator$^9$ is also of this type.
The problem is that the associated quantum algebra
includes functions which are not defined on the whole momentum
phase space,
and elements of the carrier space fail to be smooth
sections of the quantum bundle.  Indeed, written with respect to the
action-angle variables, this quantum
algebra consists of functions which are affine in angle coordinates.

We choose a different polarization
spanned by almost-Hamiltonian vector fields $\dr^k$ of angle variables.
The associated quantum algebra $\cA$ 
consists of smooth functions which are affine in action variables. 
Note that this quantization 
of the symplectic annulus $P$ (\ref{00}) is equivalent to geometric quantization
of the cotangent bundle $T^*T^m$
of the torus $T^m$ with respect to the familiar 
vertical polarization. 
As is well-known, the vertical polarization of 
a cotangent 
bundle leads to its Schr\"odinger quantization. 
We show that $\cA$ possesses a set of
nonequivalent representations 
in the separable pre-Hilbert space $\Bbb C^\infty(T^m)$
of smooth complex functions
on $T^m$. In particular, the action operators read
\mar{ci7}\beq
\wh I_k=-i\dr_k -\la_k,  \label{ci7}
\eeq
where $\la_k$ are real numbers which specify different representations of $\cA$.
By virtue of the multidimensional Fourier theorem, 
an orthonormal basis for $\Bbb C^\infty(T^m)$ consists of
functions
\mar{ci15}\beq
\psi_{(n_r)}=\exp[i(n_r\f^r)], \qquad (n_r)=(n_1,\ldots,n_m)\in\Bbb Z^m. 
\label{ci15}
\eeq
With respect to this basis, the action operators (\ref{ci7}) are 
countable diagonal matrices
\mar{ci9}\beq
\wh I_k\psi_{(n_r)}=(n_k-\la_k)\psi_{(n_r)}. \label{ci9}
\eeq
Given the representation (\ref{ci7}),
any polynomial Hamiltonian $\cH(I_k)$ of
a CIS is uniquely quantized as a Hermitian element
$\wh\cH(I_k)=\cH(\wh I_k)$ of the enveloping algebra
$\ol\cA$ of $\cA$. It has the countable time-independent spectrum
\mar{ww10}\beq
\wh \cH(I_k)\psi_{(n_r)}=E_{(n_r)}\psi_{(n_r)}, \qquad 
E_{(n_r)}=\cH(n_k-\la_k), \qquad n_k \in(n_r). \label{ww10}
\eeq
Similarly, polynomial first integrals 
are quantized.
Since $\wh I_k$ are diagonal, one can also quantize Hamiltonians 
$\cH(I_j)$ and first integrals $F(I_j)$
which are analytic functions on $\Bbb R^m$.

Note that, because geometric quantization is equivalent under 
canonical transformations, quantization of a time-dependent 
CIS with respect to action-angle variables induces its 
quantization with respect to initial variables near an
invariant manifold in the ambient momentum phase space. 
However, its Hamiltonian  
need not be represented in terms of first
integrals and need not belong to the quantum algebra $\ol\cA$
because it fails to be a scalar
under time-dependent transformations.
Moreover, this induced quantization can not be in general extended
to the whole momentum phase space because of the 
topological obstructions to
the existence of global action-angle coordinates.$^{13,14}$

For instance,
one usually mentions a harmonic oscillator as the simplest CIS
whose quantization in the action-angle variables
looks notoriously difficult.$^{15}$ However, a 
harmonic oscillator written 
relative to action-angle coordinates $(\f,I)$ is located in
the momentum phase space
$\Bbb R^2\setminus\{0\}$, but it is not the standard oscillator on
$\Bbb R^2$. Namely, there is a monomorphism, but not an isomorphism of the
Poisson algebra of smooth complex functions on $\Bbb R^2$ to that on
$\Bbb R^2\setminus\{0\}$. In particular,
the angle polarization on $\Bbb R^2\setminus\{0\}$
is not extended to $\Bbb R^2$. As a consequence, the quantum algebra associated
to this polarization 
is not extended to $\Bbb R^2$, and so is its carrier
space $\Bbb C^\infty(T^m)$. 

In conclusion, let us remark that, since Hamiltonians depend only on 
action variables and possess
time-independent countable spectra, quantum CISs look especially 
promising for 
holonomic quantum 
computation, based on driving degenerate eigenstates of a Hamiltonian over a 
parameter space.$^{16-18}$ We will construct the corresponding quantum 
control operator.
\bigskip

\noindent
{\bf II. CLASSICAL COMPLETELY INTEGRABLE SYSTEMS}
\bigskip

Recall that the configuration space of time-dependent mechanics
is a fiber bundle $Q\to \Bbb R$
over the time axis $\Bbb R$ equipped with the bundle
coordinates $(t,q^k)$, $k=1,\ldots,m$.
The corresponding momentum phase space is the vertical
cotangent bundle
$V^*Q$ of $Q\to\Bbb R$ endowed with holonomic
coordinates $(t,q^k,p_k)$.$^{19,20}$ 
The cotangent bundle
$T^*Q$, coordinated by $(q^\la,p_\la)=(t,q^k,p_0,p_k)$, 
plays a role of the homogeneous momentum phase space. It is provided 
with the canonical Liouville form $\Xi=p_\la dq^\la$, the
symplectic form
$\Om=d\Xi$, and the corresponding Poisson bracket
\mar{z7}\beq
\{f,f'\}_T =\dr^\la f\dr_\la f' -\dr_\la
f\dr^\la f', \qquad f,f'\in C^\infty(T^*Q). \label{z7}
\eeq
There is the one-dimensional trivial affine bundle
\mar{z11}\beq
\zeta:T^*Q\to V^*Q. \label{z11}
\eeq
Given its global section $h$, one can equip $T^*Q$ 
with the global bundle coordinate $r=p_0-h$.

The fiber bundle (\ref{z11}) provides
the vertical cotangent bundle $V^*Q$ with the canonical Poisson
structure $\{,\}_V$ such that
\mar{m72',2}\ben
&&\zeta^*\{f,f'\}_V=\{\zeta^*f,\zeta^*f'\}_T, \qquad \forall 
 f,f'\in C^\infty(V^*Q), \label{m72'} \\
&& \{f,f'\}_V = \dr^kf\dr_kf'-\dr_kf\dr^kf'. 
\label{m72}
\een
Its characteristic symplectic foliation coincides with
the fibration $V^*Q\to\Bbb R$.
However, the
Poisson structure (\ref{m72}) fails to set
any dynamic equation on the momentum phase space $V^*Q$
because Hamiltonian vector fields
\be
\vt_f=\dr^kf\dr_k-\dr_kf\dr^k, \qquad
\vt_f\rfloor df'=\{f,f'\}_V, \qquad f,f'\in C^\infty(V^*Q),
\ee
of functions on $V^*Q$ are vertical. 

A Hamiltonian of time-dependent mechanics is defined
as a global section
\be
h:V^*Q\to T^*Q, \qquad p_0\circ
h=-\cH(t,q^j,p_j), 
\ee
of the affine bundle $\zeta$ (\ref{z11}).$^{19,20}$ It
yields the pull-back Hamiltonian form
\mar{b4210}\beq
H=h^*\Xi= p_k dq^k -\cH dt  \label{b4210}
\eeq
on $V^*Q$. Then there exists a unique
vector field $\g_H$ on $V^*Q$ such that 
\mar{z3}\ben
&& \g_H\rfloor dt=1, \qquad \g_H\rfloor dH=0, \nonumber \\
&& \g_H=\dr_t + \dr^k\cH\dr_k- \dr_k\cH\dr^k. \label{z3}
\een
Its trajectories obey the Hamilton equation
\mar{z20}\beq
\dot q^k=\dr^k\cH, \qquad \dot p_k=-\dr_k\cH. \label{z20}
\eeq

A first integral of
the Hamilton equation (\ref{z20}) is a smooth real function $F$ 
on $V^*Q$ whose Lie derivative 
\be
\bL_{\g_H} F=\g_H\rfloor dF=\dr_tF +\{\cH,F\}_V 
\ee
along the vector field $\g_H$ (\ref{z3}) vanishes, i.e., $F$
is constant on trajectories of $\g_H$. 
A time-dependent Hamiltonian system $(V^*Q,H)$ is said to be
completely integrable
if the Hamilton equation (\ref{z20}) admits $m$ first integrals 
$F_k$ which are
in involution with respect to the Poisson bracket $\{,\}_V$ (\ref{m72})
and whose differentials $dF_k$ are linearly independent almost everywhere,
i.e., the set of points where this condition fails is
nowhere dense in 
$V^*Q$.
One can associate to this system an autonomous CIS on $T^*Q$ as follows.

Let us consider the pull-back
$\zeta^*H$
of the
Hamiltonian form
$H$ (\ref{b4210}) onto the cotangent bundle $T^*Q$. It is readily observed that 
\mar{mm16}\beq
\cH^*=\dr_t\rfloor(\Xi-\zeta^* h^*\Xi)=p_0+\cH \label{mm16}
\eeq
is a function on $T^*Q$. 
Let us regard $\cH^*$
as a Hamiltonian of an autonomous Hamiltonian system on the symplectic
manifold $(T^*Q,\Om)$. Its Hamiltonian vector field 
\mar{z5}\beq
\g_T=\dr_t -\dr_t\cH\dr^0+ \dr^k\cH\dr_k- \dr_k\cH\dr^k \label{z5}
\eeq
is projected onto the vector field $\g_H$ (\ref{z3}) on $V^*Q$ so that
\be
\zeta^*(\bL_{\g_H}f)=\{\cH^*,\zeta^*f\}_T, \qquad
\forall f\in C^\infty(V^*Q).
\ee
An immediate consequence of this relation is the following.

\begin{prop} \label{z6} \mar{z6} 
(i) Given a time-dependent CIS $(V^*Q,H; F_k)$ on $V^*Q$, the 
Hamiltonian system
$(T^*Q;\cH^*,\zeta^*F_k)$ on $T^*Q$ is completely integrable.
(ii) Let $N$ be a connected regular invariant manifold of $(V^*Q,H; F_k)$. 
Then $h(N)\subset
T^*Q$ is a connected regular invariant manifold of 
the autonomous CIS $(T^*Q;\cH^*,\zeta^*F_k)$.
\end{prop}

Hereafter, the  
vector field $\g_H$ (\ref{z3}) is assumed to be complete. In this case, 
the Hamilton equation
(\ref{z20}) admits a unique global solution through each point of the
momentum phase space $V^*Q$, and trajectories of $\g_H$
define a trivial bundle $V^*Q\to V^*_tQ$
over any fiber 
$V^*_tQ$ of $V^*Q\to \Bbb R$. Without loss of generality, one can choose
the fiber $i_0:V^*_0Q\to V^*Q$
at $t=0$.  Since $N$ is an
invariant manifold, the fibration 
\mar{ww}\beq
\xi:V^*Q\to V^*_0Q \label{ww}
\eeq
also yields the fibration of $N$ onto 
$N_0=N\cap V^*_0Q$ such that
$N\cong \Bbb R\times N_0$
is a trivial bundle. 

\bigskip

\noindent
{\bf III. TIME-DEPENDENT ACTION-ANGLE COORDINATES}
\bigskip

Let us introduce the  
action-angle coordinates around an
invariant manifold $N$ of a time-dependent CIS
 on $V^*Q$ by use of the action-angle
coordinates around the invariant manifold $h(N)$ of the
autonomous CIS on $T^*Q$ in
Proposition \ref{z6}. Since $N$ and, consequently, $h(N)$ are noncompact, we 
first prove the following.

\begin{prop} \label{z8} \mar{z8}
Let $M$ be a connected invariant manifold of an autonomous CIS
 $\{F_\la\}$, $\la=1,\ldots,n$, on a symplectic manifold
$(Z,\Om_Z)$, and let the Hamiltonian vector fields of the first integrals
$F_\al$ on $M$ be complete.
Let $U$ be a neighbourhood of $M$ such that $\{F_\la\}$ have
no critical points in $U$ and the submersion $\times F_\la: U\to \Bbb R^n$
is a trivial bundle of Lagrangian invariant manifolds
over a domain $V'\subset \Bbb R^n$. 
Then $U$ is isomorphic
to the symplectic annulus 
\mar{z10}\beq
W'=\Bbb R^{n-m}\times T^m\times V'  \label{z10}
\eeq
provided with the generalized action-angle coordinates 
\mar{z11'}\beq
(x^1,\ldots, x^{n-m},\f^{n-m+1},\ldots,\f^n,I_1,\ldots,I_n) \label{z11'}
\eeq
such that the symplectic form on $W'$ reads
\be
\Om_Z=dI_a\w dx^a +dI_k\w d\f^k,
\ee
and the first integrals $F_\la$ depend only on  
the action coordinates $(I_\al)$.
\end{prop}

\begin{proof}
In accordance with the well-known theorem,$^{5,21}$ the invariant
manifold $M$ is diffeomorphic to the product $\Bbb R^{n-m}\times T^m$,
provided with coordinates $(y^\la)=(s^a,\vf^i)$ where $\vf^i$ 
are linear functions 
of parameters $s^\la$ along the integral curves of Hamiltonian vector fields
of first integrals $F_\la$ on $M$. Let $(J_\la)$ be coordinates on
$V'$ which 
are values of first integrals $F_\la$. Since
$W'\to V'$ is a trivial bundle, $(y^\la,J_\la)$ are bundle coordinates
on the annulus $W'$ (\ref{z10}). It should be emphasized that, since
group parameters are given up to a shift, the coordinates $y^\la$ on $W'$
are determined up to a shift by functions of coordinates $J_\la$. 
Written relative to these
coordinates, the symplectic form $\Om_Z$ on $W'$ reads
\mar{ww20}\beq
\Om_Z=\Om^{\al\bt}dJ_\al\w dJ_\bt + \Om^\al_\bt dJ_\al\w dy^\bt. \label{ww20}
\eeq
 By the definition of coordinates $(y^\la)$, the
Hamiltonian vector fields $\vt_\la$ of first integrals take the
coordinate form $\vt_\la=\vt_\la^\al(J_\m)\dr_\al$ where
\mar{ww25}\beq
\vt_a=\dr_a +\vt_a^i(J_\la)\dr_i, \qquad \vt_i=\vt_i^k(J_\la)\dr_k, \label{ww25}
\eeq
and they obey the relations
\mar{ww22}\beq
\vt_\la\rfloor\Om_Z=-dJ_\la,\qquad
\Om^\al_\bt \vt^\bt_\la=\dl^\al_\la. \label{ww22}
\eeq
It follows that $\Om^\al_\bt$ is a nondegenerate matrix and 
$\vt^\al_\la=(\Om^{-1})^\al_\la$, i.e., the matrix functions $\Om^\al_\bt$
depend only on coordinates $J_\la$. In Appendix A, we obtain the desired 
coordinates 
\mar{ww21}\beq
\qquad x^a=s^a, \qquad \f^a(s^b,\vf^i,J_\la), 
\qquad I_a=J_a,  \qquad I_i(J_k). \label{ww21}
\eeq
\end{proof}

Note that, if $M$ is a compact invariant manifold, the conditions of 
Proposition \ref{z8} always hold.$^6$

Of course, the generalized action-angle coordinates (\ref{z11'}) by no 
means are unique. 
For instance, let $\cF_a$, $a=1,\ldots, n-m$ be an arbitrary smooth function
on $\Bbb R^m$. 
Let us consider the canonical coordinate transformation 
\mar{ww26}\beq
x'^a=x^a, \qquad \f'^i= \f^i + 
x^a\dr^i\cF_a(I_j), \qquad I'_a=I_a-\cF(I_j), \qquad I'_k=I_k. \label{ww26}
\eeq
Then $(x'^a,\f'^k,I'_a,I'_k)$ also generalized action-angle
coordinates on the symplectic annulus which differs from $W'$ (\ref{z10})
in another trivialization. 

Now, we apply Proposition \ref{z8} to the CISs in Proposition \ref{z6}. 

\begin{prop} \label{z13} \mar{z13}
Let $N$ be a connected regular invariant manifold of a time-dependent CIS
 $(V^*Q,H;F_k)$, and let the image $N_0$ of its 
projection $\xi$ (\ref{ww}) be compact.
Then the invariant manifold $h(N)$ of the autonomous CIS
 $(T^*Q;\cH^*,\zeta^*F_k)$ has an open
neighbourhood $U$ obeying the condition of Proposition \ref{z8}.
\end{prop}

The proof is in Appendix B. 
In accordance with Proposition \ref{z8}, the open neighbourhood $U$ of
the invariant manifold $h(N)$ in Proposition \ref{z13}
is isomorphic to the symplectic annulus
\mar{z41}\beq
W'=\Bbb R\times T^m\times V',  \label{z41}
\eeq
provided with the generalized action-angle coordinates 
$(t,\f^1,\ldots,\f^m,I_0,\ldots,I_m)$
such that 
the symplectic form on 
$W'$ reads
\be
\Om'=dI_0\w dt + dI_k\w d\f^k. 
\ee
A glance at the Hamiltonian vector field $\vt_0=\g_T$ (\ref{z5})
and the relation (\ref{ww22}) - (\ref{ww21}), shows that 
$I_0=J_0=\cH^*$ and the corresponding generalized angle coordinate 
is $x^0=t$, while the first integrals $J_k=\zeta^*F_k$
depend only on the 
action coordinates $I_i$. 

Since the action coordinates $I_i$ are independent of the coordinate
$J_0$, the symplectic annulus $W'$ (\ref{z41}) inherits the fibration (\ref{z11})
which reads
\be
\zeta: W'\ni (t,\f^i,I_0,I_i)\to 
(t,\f^i,I_i)\in W=\Bbb R\times T^m\times V. 
\ee
By the relation similar to (\ref{m72'}), the product $W$ 
is provided with the Poisson structure
\mar{ww2}\beq
\{f,f'\}_W = \dr^if\dr_if'-\dr_if\dr^if', \qquad f,f'\in C^\infty(W).
\label{ww2}
\eeq
Therefore, one can regard $W$ with coordinates $(t,\f^i,I_i)$
as the momentum phase space of the 
time-dependent CIS
in question around its invariant manifold $N$. 

It is readily observed that the Hamiltonian vector field $\g_T$ of the 
autonomous Hamiltonian
$\cH^*=I_0$ is $\g_T=\dr_t$, and so is its projection $\g_H$ (\ref{z3})
on $W$. Consequently, the Hamilton equation (\ref{z20})
with respect to the action-angle coordinates take the form
$\dot I_i=0$, $\dot\f^i=0$.
Hence, $(t,\f^i,I_i)$ are the initial date coordinates.
One can introduce such coordinates as follows. Given the fibration $\xi$ (\ref{ww}),
let us provide $N_0\times V\subset V^*_0Q$ in Proposition \ref{z13} with 
action-angle coordinates $(\ol \f^i,\ol I_i)$ for the 
CIS $\{i_0^*F_k\}$ on the symplectic leaf
$V^*_0Q$. Then, it is readily observed that $(t,\ol \f^i,\ol I_i)$ are 
time-dependent action-angle coordinates
on $W$ (\ref{z48}) such that the Hamiltonian 
$\cH(\ol I_j)$ of a time-dependent CIS relative to these coordinates vanishes,
i.e., $\cH^*=\ol I_0$. Using the canonical transformations (\ref{ww26}),
one can consider time-dependent action-angle coordinates
besides the initial date ones. 
Given a smooth function $\cH$ on $\Bbb R^m$,
let us further provide  $W$ with
the action-angle coordinates 
\be
\f^i=\ol \f^i + t\dr^i\cH(\ol I_j), \qquad
I_0=\ol I_0-\cH(\ol I_j), \qquad I_i=\ol I_i
\ee
such that $\cH(I_i)$ is the Hamiltonian. 

Thus, action-angle coordinates for a time-dependent CIS
provide a particular solution of the problem of a representation of its
Hamiltonian in terms of first integrals.$^{22,23}$ However, this representation
need not hold with respect to the initial bundle coordinates on $V^*Q$ 
because a
Hamiltonian fails to be a scalar under time-dependent
transformations.

\bigskip

\noindent
{\bf IV. QUANTUM COMPLETELY INTEGRABLE 
 SYSTEMS}
\bigskip

In order to quantize a time-dependent CIS on the
Poisson manifold $(W,\{,\}_W)$, one may follow the general procedure of
instantwise 
geometric quantization of time-dependent Hamiltonian systems in Ref. 24.
As was mentioned above, it however can be quantized
as an autonomous CIS on the symplectic annulus $(P,\Om_P)$ (\ref{00})
with respect to fixed time-dependent action-angle coordinates.

In accordance with the standard geometric quantization procedure,$^{7,8}$
since the symplectic form $\Om_P$ (\ref{ci1}) is exact, the prequantum bundle
is defined as a trivial complex line bundle $C$ over $P$. 
Since the action-angle coordinates are canonical for the symplectic
form (\ref{ci1}), the prequantum bundle $C$ need no metaplectic 
correction.
Let its trivialization
\mar{ci3}\beq
C\cong P \times \Bbb C \label{ci3}
\eeq
hold fixed. Any other trivialization leads to
equivalent quantization of $P$.
Given the associated bundle coordinates $(\f^k,I_k,c)$, $c\in\Bbb C$, on 
$C$ (\ref{ci3}),
one can treat its sections as smooth complex functions on
$P$.

The Konstant--Souriau prequantization formula associates to
each smooth real function $f\in C^\infty(P)$ on
$P$ the first order differential operator
\mar{lqq46}\beq
\wh f=-i\nabla_{\vt_f} + f \label{lqq46}
\eeq
on sections of $C$, where $\vt_f=\dr^kf\dr_k -\dr_kf\dr^k$
is the Hamiltonian vector field of $f$ and
$\nabla$ is the covariant differential with respect to a
suitable $U(1)$-principal connection on $C$. This connection
preserves the
Hermitian metric $g(c,c')=c\ol c'$ on $C$, and
its curvature form obeys the prequantization
condition $R=i\Om_P$. It reads
\mar{ci20}\beq
A=A_0 +icI_kd\f^k\ot\dr_c, \label{ci20}
\eeq
where $A_0$ is a flat $U(1)$-principal connection on $C\to
P$. The equivalence classes of flat
principal connections on $C$ are indexed
by the set $\Bbb R^m/\Bbb Z^m$ of homomorphisms of the de Rham cohomology
group
\be
H^1(P)=H^1(T^m)=\Bbb R^m
\ee
of $P$ to the cycle group $U(1)$.$^9$
We choose their representatives of the form
\be
A_0[(\la_k)]=dI_k\ot\dr^k + d\f^k\ot(\dr_k +i\la_kc\dr_c),
\qquad \la_k\in [0,1).
\ee
Then the connection (\ref{ci20}) up to gauge transformations 
reads
\mar{ci14}\beq
A[(\la_k)]=dI_k\ot\dr^k + d\f^k\ot(\dr_k +i(I_k+\la_k)c\dr_c).
  \label{ci14}
\eeq
For the sake of simplicity, we will assume that the numbers $\la_k$
in the expression(\ref{ci14}) belong to $\Bbb R$, but will bear in mind that
connections $A[(\la_k)]$ and $A[(\la'_k)]$ with $\la_k-\la'_k\in\Bbb Z$
are gauge conjugated. Given a connection (\ref{ci14}),
the prequantization operators (\ref{lqq46}) read
\mar{ci4}\beq
\wh f=-i\vt_f +(f-(I_k+\la_k)\dr^kf). \label{ci4}
\eeq

Let us choose the above mentioned angle polarization $V\pi$ which is 
the vertical tangent bundle of the fibration $\pi:P\to T^m$, and
is spanned by the vectors $\dr^k$. 
It is readily observed that the corresponding quantum algebra 
$\cA\subset C^\infty(P)$
consists of affine functions
\mar{ci13}\beq
f=a^k(\f^j)I_k + b(\f^j) \label{ci13}
\eeq
of action coordinates $I_k$. 
The carrier space of its representation by operators (\ref{ci4}) is 
defined as the
space $E$ of sections $\rho$ of the prequantum bundle $C$ of 
compact support
which obey the condition $\nabla_\vt\rho=0$ for any Hamiltonian vector field
$\vt$ subordinate to the distribution $V\pi$. This condition reads
\be
\dr_kf\dr^k\rho=0, \qquad \forall f\in C^\infty(T^m).
\ee
It follows that elements of $E$ are independent of action variables and,
consequently, fail to be of compact support, unless $\rho=0$.
This well-known problem of Schr\"odinger geometric quantization 
is solved as follows.$^{24,25}$

Fix a slice $i_T:T^m\to T^m\times V$.
Let $C_T=i^*_TC$ be the pull-back of the prequantum bundle $C$ (\ref{ci3})
over the torus $T^m$. It is a trivial complex line bundle $C_T=T^m\times\Bbb C$
provided with the pull-back Hermitian metric
$g(c,c')=c\ol c'$. Its sections are smooth complex functions on
$T^m$. Let
\be
\ol A = i^*_TA= d\f^k\ot(\dr_k +i(I_k+\la_k)c\dr_c)
\ee
be the pull-back of the connection $A$ (\ref{ci14}) onto $C_T$.
Let $\cD$ be a metalinear bundle
of complex half-forms on the torus $T^m$.
It admits the canonical lift
of any vector field $\tau$ on $T^m$, and
the corresponding
Lie derivative of its sections reads
\be
\bL_\tau=\tau^k\dr_k+\frac12\dr_k\tau^k.
\ee
Let us consider the tensor product
\mar{ci6}\beq
Y=C_T\ot\cD\to T^m. \label{ci6}
\eeq
Since the Hamiltonian vector fields
\be
\vt_f=a^k\dr_k-(I_r\dr_ka^r +\dr_kb)\dr^k
\ee
of functions $f$ (\ref{ci13}) are projectable onto $T^m$, one can
associate to each
element $f$ of the quantum algebra $\cA$ the first order
differential operator
\mar{lmp135}\beq
\wh f=(-i\ol\nabla_{\pi\vt_f} +f)\ot\id+\id\ot\bL_{\pi \vt_f}=
-ia^k\dr_k-\frac{i}{2}\dr_ka^k-a^k\la_k +b \label{lmp135}
\eeq
on sections of $Y$. A direct
computation shows
that the operators (\ref{lmp135}) obey the Dirac condition
\be
[\wh f,\wh f']=-i\wh{\{f,f'\}}. 
\ee
Sections $\rho_T$ of the quantum bundle $Y\to T^m$ (\ref{ci6})
constitute a pre-Hilbert space $E_T$ with respect to the nondegenerate
Hermitian
form
\be
\lng \rho_T|\rho'_T\rng=\left(\frac1{2\pi}\right)^m\op\int_{T^m}
\rho_T \ol \rho'_T, \qquad \rho_T,\rho'_T\in E_T.
\ee
Then it is readily observed that $\wh f$ (\ref{lmp135}) are Hermitian operators
in $E_T$. In particular, the action operators take the form (\ref{ci7}).

Of course, the above quantization depends on the choice of a
connection $A[(\la_k)]$ (\ref{ci14}) and
a metalinear bundle $\cD$. The latter need not be trivial.
If
$\cD$ is trivial, sections of the quantum bundle $Y\to T^m$ (\ref{ci6})
obey the transformation
rule
\be
\rho_T(\f^k+2\pi)=\rho_T(\f^k)
\ee
for all indices $k$. They are naturally complex smooth functions on $T^m$.
In this case, $E_T$ is the above mentioned pre-Hilbert space 
$\Bbb C^\infty(T^m)$ of complex smooth functions on $T^m$ whose basis 
consists of functions (\ref{ci15}). The action operators $\wh I$ (\ref{ci7})
with respect to this basis are represented by
countable diagonal matrices (\ref{ci9}), while functions $a(\f)$
are decomposed
into the pull-back functions $\psi_{(n_r)}$
which act on $\Bbb C^\infty(T^m)$ by multiplications
\mar{ci11}\beq
\psi_{(n_r)} \psi_{(n'_r)}=\psi_{(n_r)} 
\psi_{(n'_r)}=\psi_{(n_r+n'_r)}. \label{ci11}
\eeq

If $\cD$ is a nontrivial metalinear bundle, sections of the quantum bundle
$Y\to T^m$ (\ref{ci6}) obey the transformation
rule
\mar{ci8}\beq
\rho_T(\f^j+2\pi)=-\rho_T(\f^j)  \label{ci8}
\eeq
for some indices $j$. In this case, the orthonormal basis of the 
pre-Hilbert space
$E_T$ can be represented by double-valued complex functions
\mar{ci10}\beq
\psi_{(n_i,n_j)}=\exp[i(n_i\f^i+ (n_j+\frac12)\f^j)] \label{ci10}
\eeq
on $T^m$. They are eigenvectors
\be
\wh I_i\psi_{(n_i,n_j)}=(n_i-\la_i)\psi_{(n_i,n_j)}, \qquad
\wh I_j\psi_{(n_i,n_j)}=(n_j-\la_j +\frac12)\psi_{(n_i,n_j)}
\ee
of the operators $\wh I_k$ (\ref{ci7}), and the functions 
$a(\f)$
act on the basis (\ref{ci10}) by the above law (\ref{ci11}).
It follows that the representation of $\cA$ 
determined by the connection
$A[(\la_k)]$ (\ref{ci14}) in the space of sections
(\ref{ci8}) of a nontrivial quantum bundle $Y$ (\ref{ci6})
is equivalent to its representation determined by the connection
$A[(\la_i,\la_j-\frac12)]$ in the space $\Bbb C^\infty(T^m)$
of smooth
complex functions on $T^m$.

Therefore, one can restrict the study of representations of
the quantum algebra $\cA$ to its representations in $\Bbb C^\infty(T^m)$ 
associated
to different connections (\ref{ci14}). These representations are 
nonequivalent, unless $\la_k-\la'_k\in\Bbb Z$ for all indices $k$.

Now, in order to quantize the Poisson manifold $(W,\{,\}_W)$, one can simply 
replace functions on $T^m$ with those on $\Bbb R\times
T^m$.$^{7,24}$ Let us choose the angle polarization of $W$ 
spanned by the vectors $\dr^k$. 
The corresponding quantum algebra $\cA_W\subset C^\infty(W)$ consists of
affine functions
\mar{ww7}\beq
f=a^k(t,\f^j)I_k + b(t,\f^j)  \label{ww7}
\eeq 
of action coordinates $I_k$, represented by the operators (\ref{lmp135})
in the space $\Bbb C^\infty(\Bbb R\times T^m)$ of smooth complex functions on
$\Bbb R\times T^m$. This space is provided with the structure of the
pre-Hilbert $\Bbb C^\infty(\Bbb R)$-module with respect to the nondegenerate 
$\Bbb C^\infty(\Bbb R)$-bilinear form
\be
\lng \psi|\psi'\rng=\left(\frac1{2\pi}\right)^m\op\int_{T^m}  \psi \ol \psi', 
\qquad \psi,\psi'\in \Bbb C^\infty(\Bbb R\times T^m).  
\ee 
Its basis consists of the pull-backs onto
$\Bbb R\times T^m$ of the 
functions $\psi_{(n_r)}$ (\ref{ci15}).

Since the Poisson structure (\ref{ww2}) defines no dynamics 
on the momentum phase space $W$ (\ref{z48}), 
we should 
quantize the homogeneous momentum
phase space $W'$ (\ref{z41}) in order to
describe evolution of a quantum time-dependent CIS. Following the
general scheme in Refs. 25,26, one can provide the relevant geometric 
quantization of the symplectic annulus $(W',\Om')$. 
The corresponding quantum algebra $\cA_{W'}\subset C^\infty(W')$ consists of
affine functions
\be
f=a^\la(t,\f^j)I_\la + b(t,\f^j) 
\ee 
of action coordinates $I_\la$. It suffices to consider its subalgebra
consisting of the elements $f$ and $I_0+f$ for all $f\in \cA_W$ (\ref{ww7}).
They are represented by the operators $\wh f$ (\ref{lmp135}) and $I_0=-i\dr_t$
in the pre-Hilbert module $\Bbb C^\infty(\Bbb R\times T^m)$.
 If a Hamiltonian $\cH(I_j)$ 
of the time-dependent CIS is a polynomial
(or analytic) function in action variables, the Hamiltonian $\cH^*$ of the
associated autonomous CIS is quantized as 
\be
\wh\cH^*=-i\dr_t +\cH(\wh I_j).
\ee  
Then we obtain the Schr\"odinger equation
\be
\wh\cH^*\psi=-i\dr_t\psi +\cH(-i\dr_k -\la_k)\psi=0, \qquad 
\psi\in  \bC^\infty(\Bbb R\times T^m). 
\ee
Its solutions are the series
\be
\psi=\op\sum_{(n_r)} B_{(n_r)} \exp[-iE_{(n_r)}t]\psi_{(n_r)}, \qquad 
B_{(n_r)}\in\Bbb C,
\ee
where $E_{(n_r)}$ are the eigenvalues (\ref{ww10})
of the Hamiltonian $\wh\cH$.

In conclusion, bearing in mind applications to 
holonomic quantum computation, let us choose action-angle coordinates
such that 
a Hamiltonian
$\cH$ of a CIS is independent of action variables 
$I_a$ ($a,b,c=1,\ldots,l$). Then its eigenvalues are countably
degenerate. Let us consider the perturbed Hamiltonian
\be
\cH'=\Delta(s^\m,\f^b,I_a) +\cH(I_j), 
\ee
where the perturbation term $\Delta$ depends on the action-angle 
coordinates with the above mentioned indices 
$a,b,c,\ldots$ and on some time-dependent parameters $s^\m(t)$ by the law
\mar{02}\beq
\Delta=\La^a_\bt(s^\m,\f^b)\dr_ts^\bt I_a. \label{02}
\eeq
The Hamiltonian $\cH'$ characterizes a CIS with 
time-dependent
parameters.$^{20,26,27}$ Being affine in action variables, the perturbation
term $\Delta$ (\ref{02}) is represented by the operator
\be
\wh\Delta=-(i\La^a_\bt\dr_a +\frac{i}{2}\dr_a\La^a_\bt
 +\la_a \La^a_\bt)\dr_ts^\bt. 
\ee
Since the operators $\wh\Delta$ and $\wh\cH$ mutually commute, the
total quantum evolution operator falls into the product
\be
T\exp\left[-i\op\int_0^t\wh\cH'dt'\right]=
T\exp\left[-i\op\int_0^t\wh\cH dt'\right]\circ 
T\exp\left[-i\op\int_0^t\wh\Delta dt'\right].
\ee
The first factor in this product is the dynamic evolution operator of
the quantum CIS. 
The second one acts in the eigenspaces of the dynamic Hamiltonian 
$\wh\cH$
and reads
\mar{zz25}\ben
&& T\exp\left[\op\int_0^t\{
-\La^a_\bt(\f^b,s^\m(t'))\dr_a -\frac12\dr_a\La^a_\bt(\f^b,s^\m(t'))
+i\la_a \La^a_\bt(\f^b,s^\m(t'))\}\dr_ts^\bt dt'\right]\nonumber\\
&&\qquad =T\exp\left[\op\int_{s([0,t])}\{
-\La^a_\bt(\f^b,\si^\m)\dr_a -\frac12\dr_a\La^a_\bt(\f^b,\si^\m)
+i\la_a \La^a_\bt(\f^b,\si^\m)\} d\si^\bt\right]. \label{zz25}
\een
It is readily observed that this operator depends on the curve 
$s([0,1])\subset S$ in the parameter space $S$. One can treat it as
an operator of parallel
displacement along the curve
$s$.$^{26-28}$
For instance, if $s([0,1])$ is a loop in $S$, the operator 
(\ref{zz25}) is  
the geometric Berry factor, and it can be treated 
as a holonomy control operator.$^{26,29}$

\bigskip

\noindent
{\bf APPENDIX A}
\bigskip

In order to complete the proof of Proposition \ref{z8}, let us first
apply the relation (\ref{ww22}) to the Hamiltonian vector fields 
$\vt_a$ and $\vt_i$ (\ref{ww25}). We obtain 
\mar{ww30,1}\ben
&& \Om^a_b=\dl^a_b, \qquad \vt_a^\la\Om^i_\la=0, \label{ww30}\\
&& \vt^k_i\Om^j_k=\dl^j_i, \qquad \vt^k_i\Om^a_k=0. \label{ww31}
\een
The first of the equalities (\ref{ww31}) shows that the matrix $\Om^j_k$
is nondegenerate, and so is $\vt^k_i$. Then the second one results in $\Om^a_k=0$.

Using the well-known K\"unneth formula for the de Rham cohomology of a product,
one can easily justify that the closed form $\Om_Z$ on $W'$ (\ref{z10})
is exact. Moreover, $\Om_Z=d\Xi$ where $\Xi$ takes the form
\be
\Xi=\Xi^\al(J_\la,y^\la)dJ_\al + \Xi_i(J_\la) d\vf^i. 
\ee
Of course, $\Xi$ is determined up to an exact form. Using the fact that
components of $d\Xi=\Om_Z$ are independent of $y^\la$ and obey the equalities
(\ref{ww30}) -- (\ref{ww31}), we obtain the following.
 
(i) $\Om^a_i=-\dr_i\Xi^a +\dr^a\Xi_i=0$. It follows that $\dr_i\Xi^a$ is 
independent of $\vf$, i.e., $\Xi^a$ is affine in $\vf$ and, consequently,
is independent of $\vf$ since $\vf$ is a cyclic coordinate. Hence, 
$\dr^a\Xi_i=0$, i.e., $\Xi_i$ is a function only of coordinates $J_j$.

(ii) $\Om^k_i=-\dr_i\Xi^k +\dr^k\Xi_i$. Similarly, one shows that $\Xi^k$
is independent of $\vf$ and $\Om^k_i=\dr^k\Xi_i$, i.e., 
$\dr^k\Xi_i$ is a nondegenerate matrix.

(iii) $\Om^a_b=-\dr_b\Xi^a=\dl^a_b$. Hence, $\Xi^a=D^a(J_\la) +s^a$.

(iv) $\Om^i_b=-\dr_b\Xi^i$, i.e., $\Xi^i$ is affine in $s^a$.

In view of items (i) -- (iv), the Liouville form $\Xi$  reads
\be
\Xi=x^adJ_a + [D^i(J_\la) + B^i_a(J_\la)s^a]dJ_i + \Xi_i(J_j) d\vf^i,
\ee
where $x^a=s^a+D^a(J_\la)$. Since the matrix $\dr^k\Xi_i$ is nondegenerate,
one can introduce new coordinates $I_i=\Xi_i(J_j)$, $I_a=J_a$. We obtain
\be
\Xi=x^adI_a + [D'^i(I_\la) + B'^i_a(I_\la)s^a]dI_i + I_i d\vf^i.
\ee
Finally, put 
\be
\f^i=\vf^i-[D'^i(I_\la) + B'^i_a(I_\la)s^a]
\ee  
in order to obtain the desired coordinates (\ref{ww21}).

\bigskip

\noindent
{\bf APPENDIX B}
\bigskip

In order to prove Proposition \ref{z13}, we 
first show that 
functions $i_0^*F_k$ make up a CIS on the symplectic
leaf $(V^*_0Q,\Om_0)$ and 
$N_0$ is its invariant manifold without critical points.
Clearly, 
the functions $i_0^*F_k$ are in involution, and $N_0$ is their
connected invariant manifold. Let us show that
the set of critical points of $\{i_0^*F_k\}$ is nowhere
dense in $V^*_0Q$ and $N_0$ has none of these points.
Let $V^*_0Q$ be equipped with some coordinates $(\ol q^k,\ol p_k)$.
Then the trivial bundle $\xi$ (\ref{ww}) is provided with the bundle
coordinates $(t,\ol q^k,\ol p_k)$ which play a role of
the initial date coordinates on the momentum phase space $V^*Q$.
Written with respect to these coordinates, the first integrals
$F_k$ become time-independent. It follows that 
\mar{ww12}\beq
dF_k(y)=di_0^*F_k(\xi(y)) \label{ww12}
\eeq
for any point
$y\in V^*Q$. In particular, if $y_0\in V^*_0Q$ is a critical point
of $\{i_0^*F_k\}$, then the trajectory $\xi^{-1}(y_0)$ is a critical set
for the first integrals $\{F_k\}$. The desired statement at once
follows from this result.

Since $N_0$ is compact and regular,
there is an open neighbourhood of $N_0$
in $V^*_0Q$ isomorphic to $N_0\times V$ where $V\subset \Bbb R^m$ is
a domain, and $N_0\times \{v\}$, $v\in V$, are also invariant
manifolds in $V^*_0Q$.$^6$ Then 
\mar{z70}\beq
W''=\xi^{-1}(N_0\times V) \cong N\times V \label{z70}
\eeq
is an open neighbourhood in
$V^*Q$ of the invariant manifold $N$ foliated by
invariant manifolds $\xi^{-1}(N_0\times \{v\})$,
$v\in V$, of the time-dependent CIS on $V^*Q$. By virtue
of the equality (\ref{ww12}),
the first integrals $\{F_k\}$ have no critical points in $W''$.
For any real number $r\in(-\ve,\ve)$, let us consider a section 
\be
h_r:V^*Q\to T^*Q, \qquad p_0\circ
h_r=-\cH(t,q^j,p_j) +r,
\ee
of the affine bundle $\zeta$ (\ref{z11}). Then the images
$h_r(W'')$ of 
$W''$ (\ref{z70}) make up an open neighbourhood $U$ of
$h(N)$ in $T^*Q$. Because $\zeta(U)=W''$, the
pull-backs $\zeta^*F_k$ of first integrals $F_k$ are free from critical points
in $U$, and so is the function $\cH^*$ (\ref{mm16}). Since
the coordinate $r=p_0-h$ 
provides a trivialization of the affine bundle $\zeta$, the open neighbourhood
$U$ of $h(N)$ is diffeomorphic to the product
\be
h(W'')\times (-\ve,\ve)\cong  
h(N)\times V\times (-\ve,\ve)
\ee
which is a trivialization of the fibration
\be
\cH^*\times(\times \zeta^*F_k): U\to V\times (-\ve,\ve).
\ee

It remains to prove that the Hamiltonian vector fields of $\cH^*$ and
$\zeta^*F_k$ on $U$ are complete. It is readily observed that the
Hamiltonian vector field $\g_T$ (\ref{z5}) of $\cH^*$ is tangent to
the manifolds $h_r(W'')$, and is the image 
$\g_T=Th_r\circ \g_H\circ \zeta$
of the vector field $\g_H$ (\ref{z3}).
The latter is complete on $W''$, and so is $\g_T$ on
$U$. Similarly, the Hamiltonian vector field
\be
\g_k=-\dr_tF_k\dr^0 +\dr^iF_k\dr_i -\dr_iF_k\dr^i
\ee 
of the function $\zeta^*F_k$ on $T^*Q$ with respect to the Poisson bracket
$\{,\}_T$ (\ref{z7}) is tangent 
to the manifolds $h_r(W'')$, and is the image 
$\g_k=Th_r\circ \vt_k\circ \zeta$
of the Hamiltonian vector field $\vt_k$
of the first integral $F_k$ on $W''$ with respect to
the Poisson bracket $\{,\}_V$ (\ref{m72}). The vector fields $\vt_k$
on $W''$ are vertical relative to the fibration
$W''\to\Bbb R$, and are tangent to compact manifolds.
Therefore, they are complete, and so are the vector fields $\g_k$ on $U$.
Thus, $U$ is the desired open neighbourhood 
of the invariant manifold $h(N)$.

\end{document}